\documentclass[twoside]{article}

\usepackage{fleqn}
\usepackage{espcrc2}
\usepackage{epsfig}
\usepackage{amstex}

\def\nostrocostrutto#1\over#2{\mathrel{\mathop{\kern 0pt \rlap 
  {\raise.2ex\hbox{$#1$}}}
  \lower.9ex\hbox{\kern-.190em $#2$}}}

\newcommand{\eref}[1]{(\ref{#1})}      

\newcommand{\N}{{\mathcal N}}
\newcommand{\Nq}{{\mathcal N}_q}
\newcommand{\Ng}{{\mathcal N}_g}
\newcommand{\Ord}{{\mathcal O}}
\newcommand{\Lpm}{\Lambda}

\catcode`@=11
\newcount\@tempcntc
\def\@citex[#1]#2{\if@filesw\immediate\write\@auxout{\string\citation{#2}}\fi
  \@tempcnta\z@\@tempcntb\m@ne\def\@citea{}\@cite{\@for\@citeb:=#2\do
    {\@ifundefined
       {b@\@citeb}{\@citeo\@tempcntb\m@ne\@citea\def\@citea{,}{\bf ?}\@warning
       {Citation `\@citeb' on page \thepage \space undefined}}%
    {\setbox\z@\hbox{\global\@tempcntc0\csname b@\@citeb\endcsname\relax}%
     \ifnum\@tempcntc=\z@ \@citeo\@tempcntb\m@ne
       \@citea\def\@citea{,}\hbox{\csname b@\@citeb\endcsname}%
     \else
      \advance\@tempcntb\@ne
      \ifnum\@tempcntb=\@tempcntc
      \else\advance\@tempcntb\m@ne\@citeo
      \@tempcnta\@tempcntc\@tempcntb\@tempcntc\fi\fi}}\@citeo}{#1}}
\def\@citeo{\ifnum\@tempcnta>\@tempcntb\else\@citea\def\@citea{,}%
  \ifnum\@tempcnta=\@tempcntb\the\@tempcnta\else
   {\advance\@tempcnta\@ne\ifnum\@tempcnta=\@tempcntb \else \def\@citea{--}\fi
    \advance\@tempcnta\m@ne\the\@tempcnta\@citea\the\@tempcntb}\fi\fi}
\catcode`@=12

\begin{document}

\title{Perturbative QCD Description of Mean Jet and Particle 
Multiplicities in $e^+e^-$ annihilation}

\author{Sergio Lupia and Wolfgang Ochs\address{Max-Planck-Institut
f\"ur Physik, (Werner-Heisenberg-Institut) \\
F\"ohringer Ring 6, 80805 M\"unchen, Germany}} 

\null

{\large

\rightline{MPI-PhT/97-72}
\rightline{November 6th, 1997}
\vspace{2cm}

 \centerline{\LARGE\bf Perturbative QCD description of mean jet and} 
\vspace{0.2cm}
 \centerline{\LARGE\bf particle multiplicities in $e^+e^-$ annihilation\footnote{\normalsize
 to be published in the Proceedings of the 
XXVII International Symposium on Multiparticle Dynamics, 
Frascati, Italy, September 1997, to appear in Nucl. Phys. {\bf B} (Proc.
Suppl.)} } 

\vspace{1.0cm}

\centerline{SERGIO LUPIA and WOLFGANG OCHS}
\vspace{1.0cm}

\centerline{\it Max-Planck-Institut f\"ur Physik}
\centerline{\it (Werner-Heisenberg-Institut)}
\centerline{\it F\"ohringer Ring 6, D-80805 M\"unchen, Germany}

\vspace{3.0cm}
\centerline{\bf Abstract}
\bigskip
{ 
\baselineskip=24pt
\noindent
 A complete numerical solution of the
evolution equation for parton multiplicities in quark
and gluon jets with initial conditions at threshold is presented.
Data on both hadron and jet multiplicities in $e^+e^-$ annihilation are
well described with a common normalization, giving further support to
the picture of Local Parton Hadron Duality.
Predictions for LEP-II energies are presented. Furthermore 
we study the sensitivity to the cutoff parameter $Q_0$ and the scale of
$\alpha_s$. 
} }

\newpage

\begin{abstract}
 A complete numerical solution of the
evolution equation for parton multiplicities in quark
and gluon jets with initial conditions at threshold is presented.
Data on both hadron and jet multiplicities in $e^+e^-$ annihilation are
well described with a common normalization, giving further support to
the picture of Local Parton Hadron Duality.
Predictions for LEP-II energies are presented. Furthermore 
we study the sensitivity to the cutoff parameter $Q_0$ and the scale of
$\alpha_s$.
\end{abstract}

\maketitle

\section{INTRODUCTION}

Perturbative QCD calculations give  
 a good description of data on jet observables as far as a scale of few
GeV's is involved in the process. For instance, in  $e^+e^-$ annihilation 
the dependence of the mean jet multiplicity on the resolution parameter
$y_c$ defined in the Durham algorithm is well reproduced by QCD calculations
in absolute normalizations down to $y_c \sim 10^{-3}$, i.e. to a scale of a
few GeV's, when leading and next-to-leading terms in $\ln
y_c$ are resummed and a matching with the full two-loop result is
performed\cite{cdfw}. In the softer region 
with scales of the order of 1 GeV or less, 
deviations from purely perturbative predictions are visible and this is 
usually explained in terms of purely non perturbative hadronization effects.  
 However, following the idea of a soft hadronization 
mechanism \cite{dkmt}, a simple picture of hadronization, 
known as local parton hadron duality~\cite{lphd}, has been indeed 
formulated, where  the same perturbative description of QCD parton cascade 
is stretched down to small scales 
of a few hundred MeV for the transverse momentum cutoff $Q_0$, and then
parton predictions are directly compared to hadron distribution up to an
overall normalization factor. The interesting feature of this picture is 
that it is  surprisingly successful in describing the 
phenomenology of hadron production, as far as 
sufficiently inclusive observables, like particle
multiplicity or  inclusive energy spectra, are concerned~\cite{khozelocal,ko}. 

Even though both parton multiplicities and jet multiplicities are described
with the same type of evolution equation, 
two different sets of parameters, and in particular, two different
normalization factors, were needed so far 
to describe phenomenologically the
experimental results on the two observables. 
Here we show that both observables can be described in an unified way, if
the complete QCD master equation which describes the evolution of a parton
cascade is exactly solved. We refer to \cite{lo2,montlo} for further
details on this calculation.

\section{THEORETICAL FRAMEWORK}

The perturbative QCD picture gives a probabilistic description of 
the production process of a multiparton final state in terms of a 
parton cascade. Formally, the physical information about this process is
entirely contained in the evolution equation for the generating functional
of the multiparton final state \cite{dkmt}; in its more general
formulation, this equation correctly takes into account 
angular ordering, energy conservation 
and the running coupling at the one-loop order; the choice of the thansvere
momentum as the variable which enters in the coupling incorporate at least
part of higher order effects. The infrared cutoff $k_\perp > Q_0$ also 
provides one with a natural regularization of the infrared singularity in
the running coupling. 
By appropriate differentiation of the equation for the generating
functional, a coupled system of two  evolution equations
for the mean parton multiplicities 
$\Nq$ and $\Ng$ in quark and gluon jets 
can be derived \cite{dkmt,cdfw,lo2}: 
\begin{gather}
\frac{d\Ng (\eta)}{d\eta} = \int_{z_c}^{1-z_c} dz 
      \frac{\alpha_s(\tilde k_\perp)}{2\pi} \bigl[ \frac{1}{2} \Phi_{gg}(z) 
 \nonumber\\  
 \{\Ng(\eta+\ln z) + \Ng(\eta+\ln (1-z))-\Ng(\eta)\} \nonumber\\
    + n_f \Phi_{gq}(z) \nonumber \\
       \{ \Nq (\eta+\ln z) + \Nq (\eta+\ln (1-z))-\Ng (\eta)\} \bigr]
\nonumber \\ 
\frac{d\Nq (\eta)}{d\eta} = \int_{z_c}^{1-z_c} dz 
     \frac{\alpha_s(\tilde k_\perp)}{2\pi} \Phi_{qg}(z) \label{eveq}\\ 
\{\Ng(\eta+\ln z) + \Nq(\eta+\ln (1-z))-\Nq(\eta)\} \nonumber.
\end{gather}
The $\Phi_{AB}$ are the AP splitting functions 
for the process $A\to B$, $N_c$ and $n_f$ are the numbers of
colours and flavours respectively,  $\eta=\ln \frac{\kappa}{Q_c}$, 
where $\kappa=Q\sin \frac{\Theta}{2}$,
with $Q$ the hard scale of the process ($\sqrt{s}$ in $e^+e^-$
annihilation) and $\Theta$ the maximum  angle between the outgoing partons.
$\alpha_s(\tilde k_\perp)$ is the one-loop running coupling, given by 
$ \alpha_s(\tilde k_\perp)=2\pi/(b \ln (\tilde k_\perp/\Lpm))$
with $b=(11N_C-2n_f)/3$ and a smooth matching of the running through the 
heavy quark thresholds\cite{lo2}. 
The argument of $\alpha_s$ is chosen to be 
the ``Durham'' transverse momentum 
\begin{equation}
\tilde k_\perp  =  {\rm min}(z,1-z) \kappa
\label{kperp}
\end{equation}

The boundaries of the integral over the parton momentum fractions $z$ 
are determined by the lower cutoff $\tilde k_\perp > Q_c$ and given by 
$z_c  =  (Q_c\sqrt{2})/Q =\sqrt{2 y_c} = e^{-\eta}$.  
Since $z_c \leq \frac{1}{2}$, one finds 
$y_c\leq \frac{1}{8}$ and then $\eta \geq \ln 2$.

The coupled system of evolution equations must of course be supplemented by 
a suitable set of initial conditions; they are fixed at threshold to be: 
\begin{equation}
\Ng(\eta)=\Nq(\eta)=1 \qquad {\rm for} \quad 0\leq\eta\leq \ln 2.
\label{init}
\end{equation}

This master equation contains the correct leading and next-to-leading terms
in an $\sqrt{\alpha_s}$ expansion, but it has been obtained by using
approximations valid in the limit of soft emission (for instance, 
the definition of $\tilde k_\perp$ given in 
(\ref{kperp}) comes from the Durham definition of $k_\perp$  
in the limit of small $z$.). 
In other words, non-logarithmic terms have been neglected in the derivation
of the master equation, and these terms give a non-negligible contribution
in the region of small $\eta$ (i.e., poor resolution (large $y_c$) or low
$cms$ energy (small $Q$)). To improve the description in this region, 
the contribution of $\Ord (\alpha_s)$ in
(\ref{eveq}) has been replaced by the explicit result for $e^+e^-\to 3$ partons in
$\Ord (\alpha_s)$\cite{cdfw,lo2}: 
\begin{equation}
\N_{corr}^{e^+e^-}(y_c) =2 \N_q(y_c)-2 \N_q^{(1)}(y_c) +\N^{3-jet}(y_c).
\label{nepem}
\end{equation}
where $\N^{(1)}_{q}$ is obtained by taking the 
first iteration of (\ref{eveq}) with the initial conditions
(\ref{init})
and the  full $\Ord (\alpha_s)$ contribution is given by: 
\begin{gather}
\N^{3-jet}(y_c) =2 \int_{1/2}^1dz_1\int_{1-z_1}^{z_1} dz_2 \Theta(d_{23}-y_c) 
\nonumber\\ 
   \frac{C_F \alpha_s(k_\perp)}{2\pi} \frac{z_1^2+z_2^2}{(1-z_1)(1-z_2)}
  \label{jet3}\\
d_{23}=\min\left(\frac{z_2}{z_3},\frac{z_3}{z_2}\right)(1-z_1)>y_c
\label{d23}
\end{gather}
where $z_1$ ($z_2$) denote the quark (antiquark) and 
$z_3=2-z_1-z_2$ the gluon momentum fractions ($C_F$ = 4/3). 

\section{PHENOMENOLOGICAL ANALYSIS}

\subsection{Multiplicities in $e^+e^-$ annihilation}

\begin{figure*}[htb]
\begin{center}
\mbox{\epsfig{file=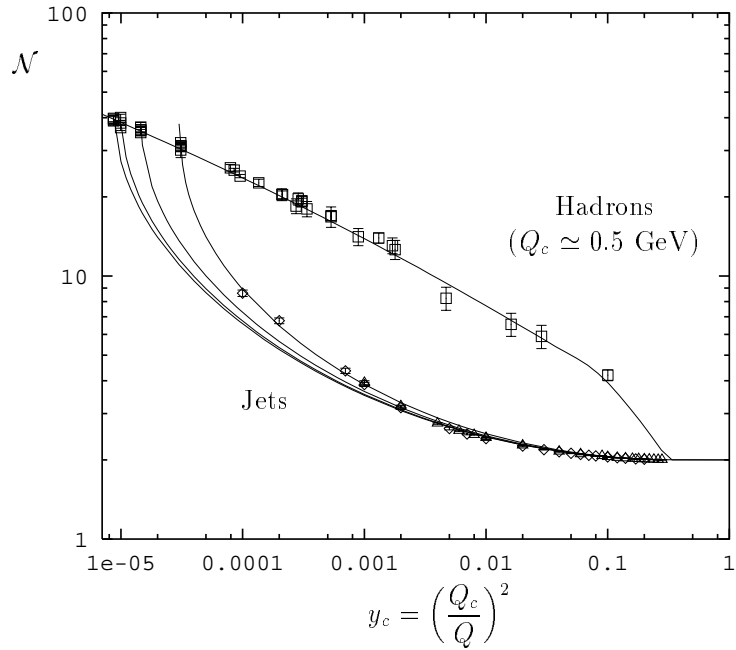,width=12cm,bbllx=2.cm,bblly=11.8cm,bburx=20.cm,bbury=23.2cm}}
\end{center}
\caption{Data on the average jet multiplicity at $Q$ = 91 GeV and
the average hadron multiplicity (assuming $\protect\N = \frac{3}{2} 
\protect\N_{ch}$)
at different $cms$
energies with $Q_c$ = 0.507 GeV as a function of $y_c$.
The solid  lines show the predictions for the hadron 
multiplicity (upper curve) and for the jet multiplicity 
at different LEP $cms$ energies (91, 133, 161 and 172 GeV) (lower curves),  
obtained by using eq.~\protect\eref{nepem} with parameters 
$K_{all}$ = 1, $\Lambda$ = 0.5 GeV and $\lambda = 0.015$. 
The right most data point for hadrons  refers to pions
only.}
\label{fig1a}
\end{figure*}

\begin{figure*}[htb]
\begin{center} 
\mbox{\epsfig{file=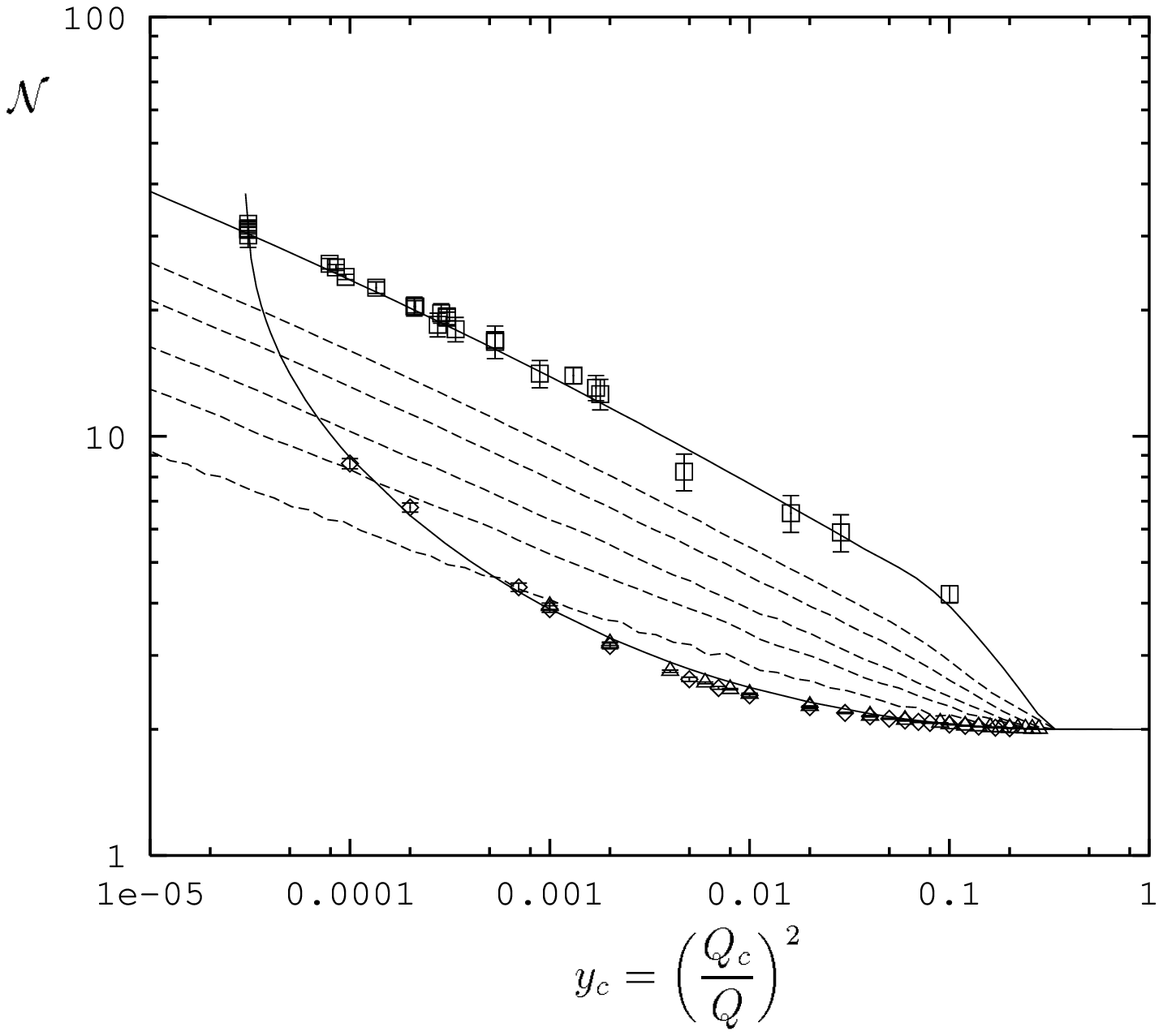,width=12cm,bbllx=2.cm,bblly=11.8cm,bburx=20.cm,bbury=23.2cm}}
\end{center}
\caption{Same data as in Fig.~1; The solid lines show predictions 
for the hadron multiplicity and for the jet multiplicity at 91 GeV 
with the same parameters as in Fig.~1; 
the dashed line show predictions for the parton  
multiplicity corresponding to the same $\Lambda$ and $K_{all}$ but 
different values of  the cutoff parameter $Q_0$= 2, 1, 0.75, 0.66, 0.55 GeV
(from bottom to top).} 
\label{fig1b}
\end{figure*}

The average jet multiplicity at $Q$ =91 GeV
as a function of the resolution parameter $y_c= Q^2_c/Q^2$, defined via  
the Durham algorithm, is shown in Figure~1\cite{lo2} (lower data
points).  Also shown in the lower part of the Figure is the theoretical
prediction in absolute normalization of the jet multiplicity at the same 
$cms$ energies of 91 GeV, as 
obtained from (\ref{nepem}). The predictions depend on the single
parameter $\Lambda$ only, so we can use the LEP-1 data on jet multiplicity to
fix the best value of the QCD scale $\Lambda$. 
The best value of $\Lambda$ turns out to be $500\pm50$ MeV, in qualitative
agreement with previous results within the MLLA framework\cite{cdfw}. 
The main improvement of our new approach in comparison to the old results is
that we can now describe experimental data down to very small values of the 
resolution parameter $y_c$; 
therefore, after having properly taken into account the effects
of energy conservation, the validity of the perturbative picture 
can be extended well below the scale of 2-3 GeV 
which was usually claimed to be the soft scale where hadronization effects
beyond perturbative QCD take place. 
In the Figure the predictions on the jet multiplicity at LEP-2 $cms$
energies are also shown. These curves could be compared with new data 
recently collected at CERN. 
After having  fixed the first parameter $\Lambda$, 
we can now ask ourselves whether it is possible to describe also 
the $cms$ energy dependence of hadron
multiplicity within the same framework. 
In this case, we can still choose two free
parameters, the hadronic scale related to the $k_\perp$ cutoff $Q_0$ and 
the normalization parameter $K_{all}$, which relates parton and hadron
multiplicities as $\N_{all}=K_{all} \N_{corr}^{e^+e^-}$. 
It turns out that a common description is indeed possible, with parameters
given by $K_{all}$ = 1 and $\lambda = \ln Q_0/\Lambda = 0.015 \pm 0.005$ 
(i.e., $Q_0 \sim$ 507 MeV).  
Figure~1 (upper data points) shows 
the data on hadron multiplicities from $\sqrt{s}$ = 1.6 (only pions) up to  
172 GeV~\cite{lo2} taken as $\N_{all}=\frac{3}{2} \N_{ch}$ 
as a function of the resolution parameter $y_c=Q^2_0/Q^2$, together with 
the perturbative predictions of (\ref{nepem}) with the aforementioned
parameters. 
It is clear from the Figure that the perturbative approach can describe
quantitatively the hadron multiplicity even at very small $cms$ energies, 
thus extending again the range of validity of the perturbative picture. 
We refer to \cite{montlo} for a more detailed comparison of our new
numerical results with previous analytical approximate solutions. 
Let us stress that a common description of both hadron and jet
multiplicities is possible with the common normalization $K_{all}=1$, 
which is a natural value, since it provides  the correct boundary conditions 
at threshold for both hadrons and jets.

It is interesting to study the dependence of our results on the 
infrared cutoff $Q_0$; one could indeed stop the parton evolution at a
larger scale of the order of 1-2 GeV  and then 
fill the gap between partonic predictions and experimental data using
purely nonperturbative hadronization models, as implemented in 
the most used Monte Carlo models. 
The perturbative predictions for the parton multiplicity with different
values of the infrared cutoff $Q_0$ are shown in Fig.~2. 
Notice that in our picture most of the particles ($\sim 3/4$) are
produced in the very last stage of the evolution for 0.5<$Q_0$<1 GeV, 
where Monte Carlo models already stopped the perturbative phase. 
A different value of $Q_0$ also changes the
energy dependence of the predicted parton multiplicity. It is indeed 
remarkable that the right energy dependence of particle multiplicity can be
reproduced with a small value of the infrared cutoff $Q_0$ and a
normalization factor $K_{all}$ = 1. A residual correlation among the two
parameters exists, so a variation of one of the two parameters within 30\%
can be accomodated by adjusting the other one. 

\begin{figure*}[htb]
\begin{center} 
\mbox{\epsfig{file=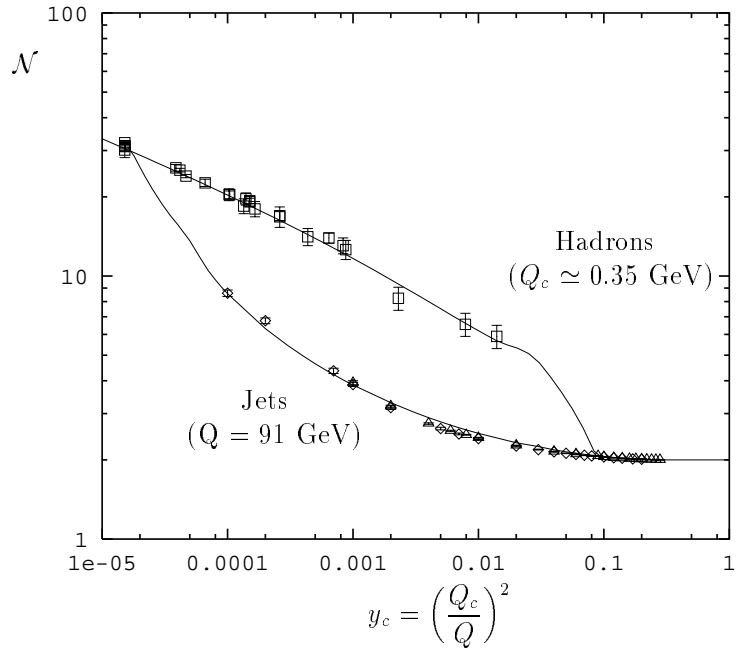,width=12cm,bbllx=2.cm,bblly=11.8cm,bburx=20.cm,bbury=23.2cm}}
\end{center}
\caption{Same as in Fig 1, but with $\Lambda$ = 0.35 GeV and a different
definition of $k_\perp$ (see text for details).}
\label{fig350}
\end{figure*}

It is important to stress that
the evolution variable $\eta=\ln \frac{\kappa}{Q_c}$ in (\ref{eveq}) 
is given by 
the ratio of the two dimensional scales which enter in the problem,
i.e., the external scale $Q$ and the resolution scale $Q_c$. Therefore, the same
evolution equation controls both the $cms$ energy dependence of the cluster
multiplicity, with clusters defined at a given resolution (where $Q$ varies
and $Q_c$ is fixed), and the
resolution dependence of the jet multiplicity at a fixed hard scale $Q$ 
(where $Q$ is fixed and $Q_c$ varies).
There is indeed only one place where the two scales enter separately and not
through their ratio, namely the expression of the running coupling
$\alpha_s$, where only $Q$ (and $\Lambda$) but not $Q_c$ appears.
Therefore, if one would switch off the running of the coupling, 
the equation would depend only on the dimensionless variable $\eta$ and
there would be no difference at all between jet and hadron multiplicities. 
It is then clear that the difference between the two multiplicities
entirely comes from the running of the coupling. Notice that this difference
goes up to a factor 10 in the region of large $y_c$ (i.e., small $cms$
energy for hadron multiplicity), and this factor is within this picture 
completely related to the increased coupling at small transverse momentum. 
This result further confirms previous results on energy
spectra\cite{lo,klo}. 
The effect of the running of the coupling is visible not only in the
difference between the two curves, in particular at large values of $y_c$,
but it is also directly visible in the strong rise of the jet multiplicity
in the high resolution region (low $y_c$) (see Fig. 2). 
Indeed, the jet multiplicity
increases roughly by a factor 3 if one lowers the resolution cutoff 
from 0.9 down to 0.5 GeV, as expected by the quick growing of the running
coupling close to the Landau singularity (which is however screened by the
$Q_c$ cutoff).  Such results are qualitatively consistent with  
expectations from the Double Log Approximation\cite{lo2}.

\subsection{Ratio of hadron multiplicities in quark and gluon jets}

The numerical solution of the coupled system of evolution 
equations (\ref{eveq}) provides one with new results on the mean
multiplicity $\N_g$ in gluon jets, and then also for 
 the ratio of hadron multiplicities in quark and gluon jets, 
$r=\N_g/\N_q$. This observable has received a lot of attention both
theoretically and experimentally. 
Theoretically, different predictions were derived for the mean parton
multiplicity in a full hemisphere of the gluon jet produced by 
 a primary $gg$ colour singlet state. 
The asymptotic value of $r$ = 9/4 has been successively reduced to smaller
values, by including next-to-leading order and even part of higher order  
correction (see \cite{lo2,montlo} for a discussion of this point). 
Experimentally, a completely inclusive
configuration similar to the theoretical predictions has recently 
been realized in the decay $Y\to gg\gamma$  by the CLEO
Collaboration \cite{cleo} and  in $e^+e^-\to 3$ jets
with nearly parallel $q$ and $\overline q$ recoiling
against the gluon by the OPAL Collaboration\cite{opalglu}.

By numerically solving the evolution equation (\ref{eveq}) 
with initial condition (\ref{init}) (without any finite order 
correction term yet), the prediction for the ratio $r$ is further reduced 
with respect to the previous approximate calculations and 
can describe the OPAL data for jets of 40 GeV very well. 
One fails, however, to reproduce the CLEO data for jets of about 5 GeV. 
This failure could be due to an important contribution from the 
fixed order correction term, not yet included so far. 

\subsection{Another definition of transverse momentum}

All results we have shown so far have been obtained by using the
Durham-inspired definition of transverse momentum given in (\ref{kperp}). 
With this method, a value of the parameter $\lambda$  
consistent with the previous finding 
 from the energy moments \cite{lo} has been found; 
however, the scale $\Lambda$ (or, equivalently, $Q_0$) has turned out 
 to be larger than the value of $\simeq$ 250 MeV used in the
description of energy spectra\cite{lphd,ko,lo}. 
In the latter case, the alternative definition of $k_\perp = z (1-z) \kappa$ 
 has been used. 
To check whether this difference plays an important role, an 
alternative calculation with this different definition of the transverse
momentum has been  performed. As shown in Fig.~3, 
a good description of experimental data is indeed 
possible also in this case, but  with a lowered scale 
 $\Lambda \sim 0.35$ GeV and the same $\lambda$ parameter. 
The remaining difference to the earlier $\Lambda$ 
comes from having taken the new scale 
$\kappa = \sqrt{2} E$  instead of $\kappa = E$ in the new calculation. 
Therefore, the new results presented here are completely
consistent with  previous studies of the energy spectra. 

\section{CONCLUSIONS}

The complete QCD evolution equations for quark and gluon jets 
has been numerically solved. 
By including also  the full $\Ord (\alpha_s)$ correction 
for $e^+e^-$ annihilation, a common description of the resolution dependence of 
jet multiplicity at LEP-1 and of the $cms$ energy dependence of 
hadron multiplicity in the whole $cms$ energy range 
has been achieved with two free parameters only 
and an overall normalization factor consistent with 1. 
 The ratio of hadron multiplicity in gluon and quark jets 
at LEP-1 is described as well. For such inclusive observables  
 the perturbative description turned out to be valid also 
in extreme kinematical domains, thus extending the
region of validity of the perturbative approach supplemented by the Local
Parton Hadron Duality picture. Further tests of this picture and, in particular, 
of its validity for soft particles and  for less
inclusive observables are certainly needed; some of them 
are discussed in \cite{khozelocal,klo}. 

\section*{AKNOWLEDGEMENTS}

One of us (S.L.) would like to 
thank Giulia Pancheri and the whole Organizing Committee  
for the nice atmosphere created at the Conference.

\end{document}